\documentclass[aps,prl,twocolumn,showpacs,superscriptaddress]{revtex4}
\usepackage{graphicx}

\begin{document}

\title{Quasiparticle coherence and the nature of the metal-insulator phase transition in Na$_x$CoO$_2$}
\author{D. Qian}
\author{L. Wray}
\author{D. Hsieh}
\affiliation{Department of Physics, Joseph Henry Laboratories of
Physics, Princeton University, Princeton, NJ 08544}
\author{D. Wu}
\author{J.L. Luo}
\author{N.L. Wang}
\affiliation{Institute of Physics, Chinese Academy of Sciences,
Beijing 100080, China}
\author{A. Fedorov}
\author{A. Kuprin}
\affiliation{Advanced Light Source, Lawrence Berkeley Laboratory,
Berkeley, Ca 94305}
\author{R.J. Cava}
\author{L. Viciu}
\affiliation{Department of Chemistry, Princeton University,
Princeton, NJ 08544}
\author{M.Z. Hasan}
\affiliation{Department of Physics, Joseph Henry Laboratories of
Physics, Princeton University, Princeton, NJ
08544}\affiliation{Princeton Center for Complex Materials, Princeton
University, Princeton, NJ 08544}
\date{\today}

\begin{abstract}

Layered cobaltates embody novel realizations of correlated quantum
matter on a spin-1/2 triangular lattice. We report a high-resolution
systematic photoemission study of the insulating cobaltates
(Na$_{1/2}$CoO$_{2}$ and K$_{1/2}$CoO$_{2}$). Observation of
single-particle gap opening and band-folding provides direct
evidence of anisotropic particle-hole instability on the Fermi
surface due to its unique topology. Kinematic overlap of the
measured Fermi surface is observed with the $\sqrt{3}$x$\sqrt{3}$
cobalt charge-order Brillouin zone near x=1/3 but not at x=1/2 where
insulating transition is actually observed. Unlike conventional
density-waves, charge-stripes or band insulators, the on-set of the
gap depends on the quasiparticle's quantum coherence which is found
to occur well below the disorder-order symmetry breaking temperature
of the crystal (the first known example of its kind).
\end{abstract}


\maketitle

Strong electron-electron interaction (Mott physics) is known to be
the origin of metal to insulator transitions as a function of doping
and temperature in many oxides\cite{1}. Such systems exhibit strong
quantum effects if electron transport occurs in lower dimensions and
the underlying band is half-filled (spin-1/2). Recently discovered
Na$_{x}$CoO$_{2}$ is the first realization of a spin-1/2 doped Mott
insulator on a triangular lattice \cite{2,3,4}. This system exhibits
not only superconductivity \cite{5} and spin-dependent thermopower
\cite{6} but also antiferromagnetism and spin-density-wave
phases\cite{2} as well as an unusual metal-insulator transition
\cite{2,3,4,7,8}. Most Mott systems exhibit insulating phases while
doped near the natural commensurate values of the underlying lattice
\cite{1}. Surprisingly, the insulating state in cobaltates is
observed near the x=1/2 doping which is not a natural commensurate
value (x = 1/3 or 2/3) of the triangular lattice. This remains an
unsolved issue to this date. In this Letter, we report the
microscopic electron dynamics in the vicinity of the metal-insulator
phase transition in Na$_x$CoO$_2$ which reveals that the insulating
state in cobaltates is due to an anisotropic particle-hole
instability on the Fermi surface resulting from its unique
momentum-space topology with respect to the Na$^+$ charge-order.
However, unlike the conventional charge-density-waves, the on-set of
the insulating gap (order parameter) is related to the energy-scale
of the emergence of quasiparticle coherence instead of the disorder
to order symmetry breaking temperature which is the first known
example of its kind.

Electron\cite{2,3}, neutron \cite{4}, infrared (IR)\cite{7},
Shubnikov-de Haas (SdH) \cite{9} and NMR measurements\cite{10} have
been used to study the insulating state. However, ARPES studies have
so far only been carried out on the metallic cobaltates\cite{11}. In
this Letter, we report a single-particle study of the cobaltate
insulator for the first time. \begin{figure}[ht] \center
\includegraphics[width=9cm]{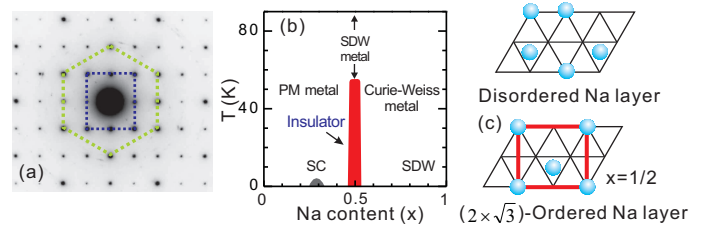}\caption{ {\bf{Phase transitions in Na$_x$CoO$_2$}} :
(a) Electron diffraction image along the [001] symmetry axis. The
rectangular supercell is observed due to the Na$^+$ charge ordering
(Wigner crystallization). (b) Phase diagram of Na$_x$CoO$_2$
\cite{2}. Metal-insulator phase transitions are observed in doping
near x=1/2 where (c) Na layer undergoes a disorder to order
transition well above the room temperature\cite{2,3,4}.}\end{figure}
High quality single crystals of Na$_{1/2}$CoO$_2$ and
K$_{1/2}$CoO$_2$ and nearby dopings were grown by the floating zone
and flux methods respectively. Na concentration near 0.5 (0.49 to
0.51) was achieved by post-growth deintercalation whereas K samples
could be naturally grown with x=0.5 (0.47 to 0.53) doping.
Insulating states with two characteristic resistivity upturns around
50K and 25K in Na-samples and 60K and 20K in K-samples
(Fig.5(a))were observed. K-based materials were used for better
surface retention of doping. No surface state was observed. High
quality of both Na and K samples allowed us to observe clear
band-folding and gap. Spectroscopic measurements were performed at
the Adv. Light Source. The data were collected with He I (21.2eV),
30 eV or 60 eV photons with better than 8, 12 or 10 meV energy
resolution and an angular resolution better than 2$\%$ of the
Brillouin zone at Beamlines 12.0.1 and 10.0.1 using Scienta
analyzers with chamber pressures better than 4$ \times$10$^{-11}$
torr.

\begin{figure}[ht] \center
\includegraphics[width=9cm]{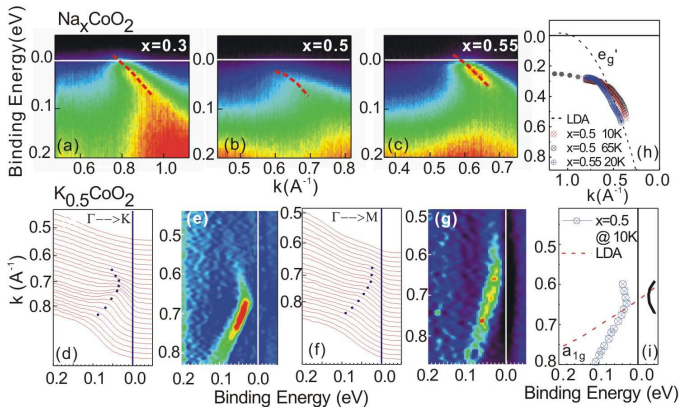}\caption{{\bf{Quasiparticle
Dynamics}}: Single-particle removal spectra of samples with doping
(a) x=0.3, (b) x=0.5 and (c) x=0.55. (d) and (e) show spectra and
color images (SDI) along $\Gamma\rightarrow$K while (f) and (g) are
spectra measured along $\Gamma\rightarrow$M. (h) Doping and
temperature dependence of e$_g$' band dispersions in the vicinity of
x=0.5 which show no changes. (i) Experimental band dispersion is
compared with LDA theory (dotted line)\cite{12}. The measured band
is less dispersive. It exhibits folding and a gap opens at the Fermi
level.}
\end{figure}

No measurable sign of sample charging was observed down to 10K. No
appreciable loss of Na-\textit{2p} (or K-\textit{3p}) peak intensity
was observed within the probe-depth of ARPES and within a few hours
of cleaving in UHV. Such care has been quite crucial in observing
the energy gap since a change of doping ($>$3$\%$) or aging turns
the sample into a metal with a large Fermi surface (FS). Measured FS
area further provided an internal check for changes in effective
doping on the surface. Rotating analyzers allowed us to polarization
select the a$_{1g}$-band states (under P-geometry) and e$_{g'}$-band
states (S-geometry) distinctively.

Fig.-2(a-c) shows the quasiparticle dispersion as the system
approaches the insulating phase from either side of the doping phase
diagram (namely x=0.3, x=0.5 and x=0.55) within 200meV of the
electron binding energies at 20K. Only in x=0.5 samples is spectral
weight suppression observed at low temperature. The gap behavior is
seen in the energy distribution curves in Fig-2(d-g). The x=0.5
samples show a clear pull-back of the quasiparticle along the high
symmetry cut $\Gamma\rightarrow$K. A spectral gap is also observed
along $\Gamma\rightarrow$M but the folding behavior is less
prominent. These quasiparticles emerge from the a$_{1g}$-like
states. We have carefully monitored changes at the e$_{g'}$ bands
under S-geometry as they are predicted to be involved in the
ordering transition\cite{13,14}. The e$_{g'}$ band is found to be
fully gapped and hardly showed any temperature or doping dependence
(Fig.-2(h)). Therefore we conclude that the insulating gap opens by
the destruction of the \textit{a$_{1g}$} Fermi surface only.

\begin{figure}[t] \center
\includegraphics[width=9.0cm]{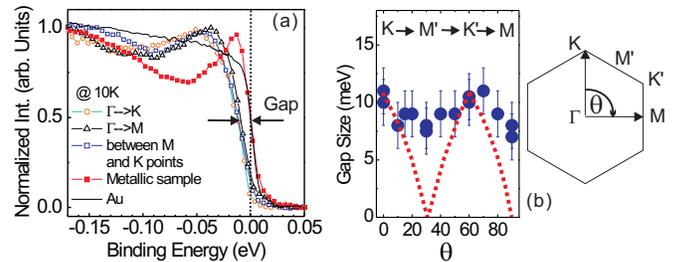}\caption{{\bf{Gap Anisotropy:}} (a) High energy-resolution
(k-integrated on 4$\%$ BZ) spectra of Na$_{0.5}$CoO$_2$ measured at
10K near the folding points along different k-space cuts. Comparison
of the spectra with Fermi-edge of Au and metallic Na$_{0.55}$CoO$_2$
demonstrates the gap in Na$_{0.5}$CoO$_2$. (b) k-dependence of
quasiparticle (leading edge) gap in K$_{0.5}$CoO$_2$ measured at
10K. Red dots are expected k-dependence of gap in case of nesting
occurring \textit{only} along $\Gamma\rightarrow$K/K' (domain mixing
effect) line. The k-space angle $\theta$ is defined from
$\Gamma\rightarrow$K to $\Gamma\rightarrow$M line.}
\end{figure}

Fig. 3(a) shows integrated (momentum window of 0.06 \AA$^{-1}$ 4$\%$
of the BZ) low energy spectra near the folding point along
$\Gamma\rightarrow$M, $\Gamma\rightarrow$K and cuts in between. The
data are compared to a spectrum from a nearby metallic sample
(x=0.55). Quasiparticles are seen to be gapped in all high symmetry
momentum space cuts. The gap energy $\Delta$ varies from 6 to 11
meV. Angular distribution of gap is plotted in Fig-3(b). Gap tends
to be larger along $\Gamma\rightarrow$K cuts.

A hexagonal Fermi surface is observed in nearby dopings (x=0.3 or
0.55) at low temperatures but no band-folding is seen. In x=0.5 the
hexagonal Fermi surface is found to be destroyed at low temperatures
by the band folding. Fig-4(a) shows the n(k) map generated with an
integration window of 20 meV ($>$ gap size). This shows the topology
of the "Fermi surface" around the onset of gap opening. This "Fermi
surface" retains hexagonal character except that its volume has
changed according to doping. The size of the measured FS at x=1/2
doping is found to be special when Na$^+$ superlattice is
considered. In reciprocal space, the primary vector G in hexagonal
cell is 2.6 \AA$^{-1}$. This is much larger than any possible
2k$_{f}$ in any doping over the phase diagram. However, considering
the Na$^+$ lattice (2 x $\sqrt{3}$ supercell BZ), the shortest
vectors Gs are 1.1 and 1.3 A$^{-1}$ which is close to the measured
value of \textit{average} 2k$_{f}$ for doping in the neighborhood of
0.5. Our results also show that the $\sqrt{3}$x$\sqrt{3}$ long-range
charge order BZ \cite{8} has no overlap with x=0.5 FS (Fig.-4(c))
hence likely not related to the insulating behavior at this
particular doping. The Na superstructure imposes a two-fold symmetry
on the FS. The Na$^+$ ionic potential likely enhances nesting
conditions ($\overrightarrow{Q}_1$$||$$\overrightarrow{b'}_1$ ,
$\overrightarrow{Q}_2$) along the $\Gamma\rightarrow$K/K' cuts
(Fig.-4(d)). This should then be reflected in the gap and folding
behavior. We also consider the domains of the supercell. With a
fixed hexagonal matrix there are 3 possible domains for the Na$^+$
supercell. The domains mix K and K' as well as M and M' \textit{but
not} M with K or M' with K'. Therefore, even in case of complete
domain mixing one would expect clear, distinct and different folding
and gap behaviors to survive along $\Gamma\rightarrow$M/M' and
$\Gamma\rightarrow$K/K' consistent with our observation. In a
\textit{conventional} density wave picture \cite{15,16} gap size is
large along the longest straight sections of the FS. Nesting
possibility along $\Gamma\rightarrow$K/K' suggests that gap would be
larger and band will exhibit stronger folding behavior since ARPES
intensity of folding is proportional to nesting strength \cite{16}.
It is likely that the effect of the sodium supercell is to flatten
out the FS also along $\Gamma\rightarrow$M which then leads to a
gap. A weaker gap would be expected in such a scenario compared to
what was actually observed along M. We note that Na$_{1/2}$CoO$_2$
is not a conventional CDW material like NbSe$_3$ (band metal) is and
other effects (finite correlation gap etc.) are likely at play near
the M-point.

\begin{figure}[ht] \center
\includegraphics[width=7.5cm]{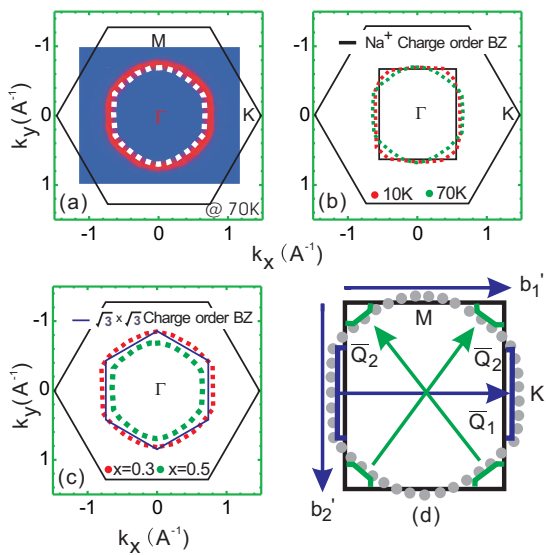}\caption{{\bf{Particle-hole instabilities on the Fermi surface:}}
Fermi surface reconstruction: Na$^+$ Charge order vs.
$\sqrt{3}$x$\sqrt{3}$ Co-Charge order (a) Momentum-distribution of
quasiparticles (2-color image), n(k), within 20 meV of Fermi energy.
The white dots reflect the FS topology. (b) Temperature dependence
(T=10K red dots, 70K green dots) of the "FS topology" in x=1/2
samples. The solid black rectangle is the BZ of Na-supercell.
Measured FS topology almost coincides with the Na-supercell. (c) The
$\sqrt{3}$x$\sqrt{3}$ Co-Charge order BZ (solid hexagon) has no
overlap with x=0.5 FS (green dots) but coincides with x=0.3
(superconducting doping, red dots) FS. (d) Gray dots are the
experimental FS. Green and blue lines are reconstructed FS contours
based on experimental folding behavior which is consistent with the
geometry of the Na supercell defined by $\vec{b'}_1$ and
$\vec{b'}_2$.}
\end{figure}

\begin{figure}[ht] \center
\includegraphics[width=6.5cm]{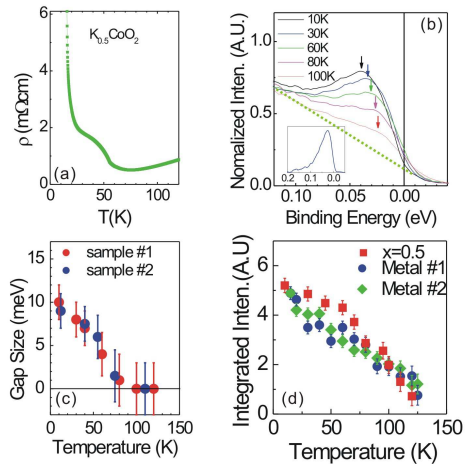}\caption{{\bf{Gap amplitude and Quasiparticle coherence}} : Temperature dependence
of (a) in-plane \textit{dc}-resistivity, (b) quasiparticle signal
along $\Gamma\rightarrow$K cut. Inset shows the quasiparticle after
linear background subtraction. (c) Evolution of leading-edge gap
with temperature. (d) The integrated spectral weight of
quasiparticle (coherence) in insulating doping (x=0.5, red dots) is
compared with the metallic dopings (x=0.6-0.7, blue and green
points) as a function of temperature. Around 150K spectral coherence
(integrated peak intensity) vanishes.}
\end{figure}

Fig.-5(b) shows the temperature dependence of quasiparticle spectral
weight. Quasiparticles get sharper and the energy gap increases as
the temperature is lowered. Fig.-5(c) shows the temperature
dependence of the charge gap. Temperature dependence of gap viewed
as a BCS-like order parameter ($\Delta$(T)) of the transition
suggests that its amplitude vanishes beyond 75-80K. This is lower
than the SDW transition ($\sim$ 88K) and suggests the gap to be
associated with the insulating transitions \cite{2}. Upon further
raising the temperature, both the gap size and the spectral weight
decrease (beyond thermal broadening). However, the spectral weight
does not completely vanish even after the gap has disappeared. Such
decrease with rising temperatures is also seen in nearby metallic
dopings (Fig-5(d)). ARPES studies have shown that in the coherent
transport regime quasiparticles gain significant weight due to
enhanced or effective out-of-plane coupling or c-axis
coherence\cite{17}.

Our observation of quasiparticle weight mainly below 150K in x=1/2
insulators suggests that even if Na/K layer is ordered well above
the room temperature ($\sim$ 350K) the cobalt states are affected by
the Na-ion potential after quasiparticles form or gain weight.
Therefore, if some approximate nesting is to be operative it can
take place only at a similar or lower temperature scale following
the growth of quasiparticles. We observe such correlation between
the charge gap amplitude and the quasiparticle coherence in
insulating cobaltates (Fig.-5). Similar behavior is seen in cuprate
superconductors (SC) where SC-gap opens shortly after the onset of
(coherent) quasiparticles which has been ascribed to be due to
out-of-plane coupling \cite{18}. In case of cobaltates, instead, it
leads to a charge-gap opening as seen in our data which is a
\textit{competing instability} with superconductivity. This also
argues that the gap we observe is not as one in a band insulator
(Slater-type) by the electron count in the new unit cell otherwise
it would have appeared at the onset of Na-order and "FS" would have
deformed to exactly coincide with the supercell BZ boundary leading
to a zero count as in a particle number vanishing phase transitions.
This does not happen even down to 10K, so there is no adiabatic
continuity to a Slater band insulator phase. It is possible that one
of the nesting/modulation vectors ($\overrightarrow{Q}_1$) become
operative (longer straight section hence stronger coupling, stronger
folding) around 60K (first upturn in resistivity (Fig.5(a)) which
nests a good part of the FS but not completely whereas at lower
temperatures near 20K both vectors ($\overrightarrow{Q}_1$,
$\overrightarrow{Q}_2$, possibly also along M) are in effect gapping
most of the FS thus leading to resistivity divergence. Such behavior
is consistent with NQR data \cite{10}.

Density wave modulations in matter form via several different
mechanisms. Singularity in the density of states helps realize a
density wave state\cite{19}. In a triangular system such
singularities in the DOS are expected in case of large and positive
\textit{t}\cite{20}. However, K$_{1/2}$CoO$_{2}$ is in the opposite
limit namely with a negative sign of hopping (\textit{t} $<$ 0,
hole-like band) with a small \textit{t$_{eff}$} ($\sim$ 16 meV
$\sim$ bandwidth(0.15 eV)/9). A more ubiquitous mechanism for
developing a density wave modulation is via FS instability. This is
the case for NbSe$_3$ and other 1-D metals that exhibit conventional
behavior\cite{21}. In most two dimensional materials (2H-NbSe$_2$
etc.\cite{22}) nesting is never complete and FS is partially gapped
and the system remains fairly metallic (no insulating divergence).
Given the transition temperature ($\sim$ 60K) in cobaltates, the gap
size (10 meV) is rather soft compared to many other systems where
$\Delta$ $>>$ k$_B$T$_c$. Gap closes much faster in cobaltates than
it is seen in most CDW systems \cite{21,22}. The coherence length of
the density wave modulation in cobaltate, based on our data, (v$_f$
$\sim$ 0.5 eV.\AA  (Fig.2) : $\xi$ $\sim$ $\hbar$v$_f$/$\Delta$
$\sim$ 10$^2$ $\AA$  is less than the correlation length of Na
charge-order \cite{4} consistent with a density wave induced on the
Co plane. Another common mechanism for generating a spectral gap is
via valence charge order as argued in \cite{2,3,4}. Many oxides
indeed exhibit charge-ordering (stripes) due to electron-electron
and electron-lattice interactions \cite{1}. Such order typically
leads to gap often known as weak correlation gap or pseudogap
\cite{1}. No overlap (intersection) of our measured Fermi Surface is
observed with the $\sqrt{3}$x$\sqrt{3}$ cobalt \textit{long-range}
charge-order Brillouin zone. However, a combination of FS
instability and correlation effects can lead to a \textit{weakly
k-dependent} (pseudo-)gap or correlation gap in manganites in the
case of \textit{fluctuating} charge-order\cite{23}. The lack of
valence disproportion occuring at the transition \cite{10} has been
used to argue for a fluctuating mechanism via intersite Coulomb (V)
which likely sets the gap amplitude\cite{24}. This can lead to a
finite correlation gap (then the weak k-modulation of the gap
reflects the fine FS topological instability effects) in cobaltates.
Irrespective of the mechanism of spectral gapping, our results are
consistent with IR \cite{7} and SdH measurements\cite{9} where it is
argued that almost \textit{entire} FS vanishes at the transition
except extremely small left-over pockets (0.25$\%$ of the hex-BZ).
Such a scenario is consistent with gap and band-folding we observe.
The reconstructed corners (Fig-4(d)) are not likely resolved due to
finite resolution and weak folding strengths (SdH \cite{9} does not
report the exact k-space locations of the pockets hence a detailed
comparison is not possible with the dispersion folding (E vs. k) and
gap we report by ARPES).

In conclusion, our results suggest that the coupling of
quasiparticle states with the crystallized Na layer indeed leads to
an intrinsic quantum many-body (correlated) insulating-like
low-energy state with weakly k-modulated gap ($\sim$ 6-11 meV) via
particle-hole instability on the Fermi surface. Kinematic overlap of
the measured Fermi surface is observed with the
$\sqrt{3}$x$\sqrt{3}$ cobalt charge-order Brillouin zone near x=1/3
but not at x=1/2. The gap opening and quasiparticle signals
(coherence) set in at a similar temperature scale which is much
smaller than the Na/K charge-order scale and exhibit spectral
redistributions over large energy scales in doping and temperature.
A comprehensive many-body theory requires to go beyond a
conventional one and needs to consider the quasiparticle coherence
effects, selective coupling to the superpotential as well as
longer-range Coulomb correlations in driving this novel phase
transition.

We gratefully acknowledge D.A. Huse, D.H. Lee, M. Lee, P.A. Lee,
N.P. Ong, P. Phillips and S. Shastry for discussions. This work is
partially supported through NSF-MRSEC (DMR-0213706) grant and the
DOE, grant DE-FG02-05ER46200.

\end{document}